\documentclass[12pt]{article}
\usepackage{axodraw}
\usepackage{epsfig}
\usepackage{amssymb}
\usepackage{latexsym}
\usepackage{amsmath}
\usepackage{color}
\usepackage{graphics}
\usepackage{subfigure}

\newcommand{\lsim}{\raisebox{-0.13cm}{~\shortstack{$<$ \\[-0.07cm] $\sim$}}~}
\newcommand{\gsim}{\raisebox{-0.13cm}{~\shortstack{$>$ \\[-0.07cm] $\sim$}}~}
\newcommand{\eqa} {\begin{eqnarray} }
\newcommand{\eqe} {\end{eqnarray}}
\newcommand{\beq} {\begin{equation}}
\newcommand{\eeq} {\end{equation}}

\begin{document}
\pagestyle{empty}
\begin{flushright}
September 2009
\end{flushright}
\begin{center}
{\large\sc {\bf Rapidity Gap Events in Squark Pair Production at the LHC}} 

\vspace{1cm}
{\sc Sascha Bornhauser$^{1}$, Manuel Drees$^{2,3}$, Herbert
  K. Dreiner$^2$} and {\sc Jong Soo Kim$^4$}  

\vspace*{5mm}
{}$^1${\it Department of Physics \& Astronomy, University of New
  Mexico, 800 Yale Blvd NE Albuquerque, NM 87131, USA} \\ 
{}$^2${\it Physikalisches Institut and Bethe Center for Theoretical
  Physics, Universit\"at Bonn, Nussallee 12, D53115 Bonn, Germany} \\ 
{}$^3${\it KIAS, School of Physics, Seoul 130--012, Korea}\\
{}$^4${\it  Institut f\"ur Physik, Technische Universit\"at Dortmund,
  D-44221 Dortmund, Germany } 
\end{center}
\vspace*{1cm}
\begin{abstract}
  
  The exchange of electroweak gauginos in the $t-$ or $u-$channel allows
  squark pair production at hadron colliders without color exchange between
  the squarks. This can give rise to events where little or no energy is
  deposited in the detector between the squark decay products.  We discuss the
  potential for detection of such rapidity gap events at the Large Hadron
  Collider (LHC). Our numerical analysis is divided into two parts. First, we
  evaluate in a simplified framework the rapidity gap signal at the parton
  level. The second part covers an analysis with full event simulation using
  PYTHIA as well as Herwig++, but without detector simulation. We analyze the
  transverse energy deposited between the jets from squark decay, as well as
  the probability of finding a third jet in between the two hardest jets. For
  the mSUGRA benchmark point SPS1a we find statistically significant
  evidence for a color singlet exchange contribution. The systematical
  differences between current versions of PYTHIA and HERWIG++ are larger than
  the physical effect from color singlet exchange; however, these systematic
  differences could be reduced by tuning both Monte Carlo generators on normal
  QCD di--jet data.

\end{abstract}

\newpage
\setcounter{page}{1}

\pagestyle{plain}
\section{Introduction}

One of the main objectives of the Large Hadron Collider (LHC) is the
search for supersymmetric (SUSY) particles \cite{susyrev}. In the
energy range of the LHC we expect squark pair production to be one of
the most important channels for the production of superparticles
\cite{qcdlo}. Since squark pairs are produced at lowest order QCD, the
production cross sections are of the order of the strong coupling
strength squared ${\cal O}(\alpha_s^2)$. Furthermore, in some
contributing subprocesses, e.g. $ud\rightarrow\tilde{u}\tilde{d}$,
both quarks in the initial state are valence quarks. As a result,
even heavy squarks have a reasonable production cross section because
the parton distribution functions of valence quarks fall off most
slowly of all partons for large Bjorken$-x$.

Squark pair production also includes contributions with electroweak (EW)
exchange particles at tree \cite{oldew,bddk1} or one-loop \cite{nlo_ew} level.
The tree--level EW contributions can change the production cross section by up
to $50\%$ \cite{bddk1}. Moreover, EW gaugino exchange in the $t-$ or
$u-$channel gives rise to events with no color connection between the produced
squarks. QCD radiation then preferentially takes place in the phase space
region between the respective color connected initial quark and final squark,
not between the two outgoing squarks. If the rapidity region between both
squarks is indeed free of QCD radiation it is called a ``rapidity gap''. The
situation is different for the lowest order QCD contribution. The final
squarks are color--connected and radiation into the region between them is
expected. This difference might allow to isolate events with electroweak
gaugino (color singlet) exchange, which could e.g. lead to new methods to
determine their masses and couplings.

The above discussion describes a single partonic reaction producing stable
squarks. In reality, the squarks will decay. Even if we assume that each
squark decays into a single jet (and a neutralino or chargino, which may decay
into leptons and the lightest neutralino, which we assume to be the lightest
superparticle [LSP]), the rapidity distribution of these jets will differ
from that of the squarks. Squark decay also leads to additional parton showers
from final state radiation. Moreover, the underlying event produced by the
beam remnants and their interactions can also deposit energy in the gap.
Finally, rapidity gap events can also occur in pure SUSY QCD processes, which
have to be considered as background in this context.

There has been quite a lot of discussion of rapidity gaps in the Standard
Model (SM), both as a possibility to probe novel features of QCD in $e^+e^-$
annihilation \cite{eejets}, deep--inelastic scattering \cite{dis} and purely
hadronic interactions \cite{chehime,enberg,ppjets}, and as a means to enhance
the signal for Higgs production from $W^+W^-$ and $ZZ$ fusion at hadron
colliders \cite{Dokshitzer:1991he,Fletcher:1993ij,higgs}. Much of this work
was triggered by the observation of true gap events in early Tevatron data
\cite{tev_gap}. However, we are not aware of any discussion of rapidity gap
events in the production of strongly interacting superparticles. We will show
that color singlet exchange can indeed lead to detectable differences in the
final state characteristics of squark pair events even after including squark
decay, hadronization, and the underlying event. However, in order to fully
exploit this potential, semi-- and non--perturbative features of the strong
interactions have to be better understood, e.g. by analyzing ordinary QCD
di--jet events.

This paper is organized as follows. In Sec.~2 we discuss the rapidity
gap signal in squark pair production at the parton level,
i.e. ignoring the underlying event and keeping the squarks stable. In
Sec.~3 we discuss our numerical results for a full simulation. The
possibility to tune Monte Carlo generators with the help of SM QCD
processes in order to reduce systematic theoretical uncertainties is
discussed in Sec.~4. The final Section contains a short summary and
concluding discussion.


\section{Rapidity Gap Events with Stable Squarks}

In this Section we explore the physics of rapidity gap events under
the simplifying assumption that squarks are stable. This serves two
purposes. First, it allows to simply describe the physical reason why
QCD radiation into a large rapidity region might be suppressed,
leading to the formation of a gap
\cite{Bjorken:1992er,DelDuca:1993pq}; this is described in the first
Subsection. Secondly, it facilitates a first comparison of parton
shower based event generator programs to each other, and to a
parton--level program using exact leading--order matrix
elements. These comparisons are presented in the second Subsection. In
the third Subsection we discuss interference between color singlet and
non--singlet exchange.

\subsection{The Basic Argument}

\begin{figure}[t!]
  \begin{center}
    \subfigure[CNS exchange  ]{\label{ncsex}
    \includegraphics[scale=1.0]{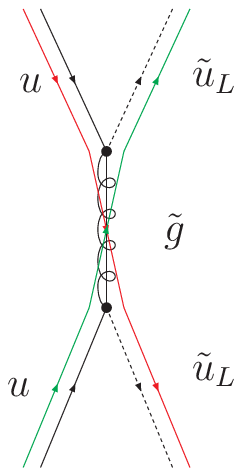}\hspace{2cm}}
    \subfigure[ CS exchange]{\label{csex}
    \includegraphics[scale=1.0]{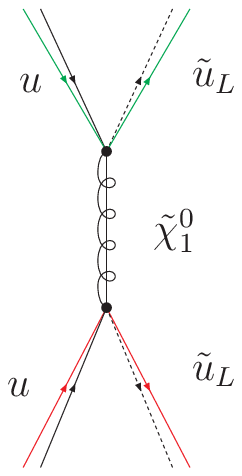}\hspace{2cm}} \\
    \subfigure[non--rapidity gap event]{\label{nrge}
{
\unitlength0.5cm
\begin{picture}(12,12)
\thicklines
\put(5,8){\color{green}\circle{6}}
\thinlines
\put(7,4){\color{red}\circle{6}}
\thinlines
\put(2,6){\color{red}\vector(1,0){4}} 
\thicklines
\put(10,6){\color{green}\vector(-1,0){4}}
\thicklines
\put(6,6){\color{green}\vector(3,2){3.4}}
\thinlines
\put(6,6){\color{red}\vector(-3,-2){3.4}}
\put(6.7,6.0){\oval(1,1)[tr]}
\put(0.5,5.5){\makebox(1,1){$\scriptstyle{u,\, p_1}$}}
\put(10.5,5.5){\makebox(1,1){$\scriptstyle{u,\, p_2}$}}
\put(10.25,8){\makebox(1,1){$\scriptstyle{\tilde u_L,\, k_1}$}}
\put(1.5,3){\makebox(1,1){$\scriptstyle{\tilde u_L,\, k_2}$}}
\put(7.5,6.0){\makebox(1,1){$\scriptstyle{\Theta_{CMS}}$}}
\end{picture}
}}
   \subfigure[rapidity--gap event]{\label{rge} {
\unitlength0.5cm
\begin{picture}(12,12)
\thicklines
\put(9.5,7){\color{green}\circle{1}}
\thinlines
\put(3,5){\color{red}\circle{1}}
\thicklines
\put(2,6){\color{green}\vector(1,0){4}} 
\thinlines
\put(10,6){\color{red}\vector(-1,0){4}}
\thicklines
\put(6,6){\color{green}\vector(3,2){3.4}}
\thinlines
\put(6,6){\color{red}\vector(-3,-2){3.4}}
\put(6.7,6.0){\oval(1,1)[tr]}
\put(0.5,5.5){\makebox(1,1){$\scriptstyle{u,\, p_1}$}}
\put(10.5,5.5){\makebox(1,1){$\scriptstyle{u,\, p_2}$}}
\put(10.25,8){\makebox(1,1){$\scriptstyle{\tilde u_L,\, k_1}$}}
\put(1.5,3){\makebox(1,1){$\scriptstyle{\tilde u_L,\, k_2}$}}
\put(7.5,6.0){\makebox(1,1){$\scriptstyle{\Theta_{CMS}}$}}
\end{picture}
} }
  \end{center}
  \begin{center}
    \parbox{16.0cm}{\caption[b]{The two Feynman diagrams show
        $t-$channel $2 \rightarrow 2$ scattering for color
        non--singlet (CNS) and color singlet (CS) exchange. The
        colored lines denote the color flow between the incoming
        quarks and outgoing squarks. The two lower diagrams indicate
        the color flow of the green (thick line) and red (thin line)
        color charge in the center of mass system; the circles lie in
        the rapidity ranges which are filled up by the gluon radiation
        off the scattered color charges. }}
  \label{fourplots}
  \end{center}
\end{figure}

Consider the parton level squark pair production process $u(p_1) u(p_2)
\rightarrow \tilde u_L(k_1) \tilde u_L(k_2)$. It can proceed through color
non--singlet (gluino) exchange, Fig.~\ref{ncsex}, as well as via color singlet
(neutralino) exchange, Fig.~\ref{csex}: in both cases both $t-$ and
$u-$channel diagrams contribute.

The pattern of gluon radiation in these reactions, which may or may not lead
to a rapidity gap, can be explained using the picture of an accelerated color
charge \cite{Fletcher:1993ij}. Figs.~\ref{nrge},(d) sketch the momentum as
well as color flow for these reactions in the center--of--mass system (CMS).
For a $t-$channel color non--singlet (CNS) exchange process the green color
charge of Fig.~\ref{ncsex} has to scatter from the momentum direction $p_2$ to
$k_1$, i.e. over an angle $\pi - \Theta_{CMS}$, with small $\Theta_{CMS}$
being preferred dynamically. As a result, we expect ``bremsstrahlung'' gluons
to be emitted by the ``green'' charge over most of the angular, or rapidity,
region, as indicated by the large green circle. The ``red'' charge is also
scattered by an angle $\pi - \Theta_{CMS}$. For the dynamically preferred case
$\Theta_{CMS} < \pi/2$ we therefore expect the entire rapidity region to be
filled with soft gluon radiation; no ``gap'' arises. This also holds for
squared $u-$channel color non--singlet exchange. Here the ``green'' color
flows from $p_2$ to $k_2$, but $\Theta_{CMS}$ close to $\pi$ is preferred
dynamically.

Now consider squared color singlet (CS) $t-$channel contribution, as sketched
in Fig.~\ref{rge}. In this case the ``green'' color is accelerated from
direction $p_2$ to direction $k_2$, i.e. it is scattered by $\Theta_{CMS}$,
with small $\Theta_{CMS}$ being dynamically preferred. The resulting
bremsstrahlung will mainly populate the region indicated by the small green
circle. Similarly, bremsstrahlung from the ``red'' color will mainly populate
the region indicated by the small red circle. Note that for $\Theta_{CMS} <
\pi/2$, {\em little or no soft bremsstrahlung is expected to occur in the
  region between the two squarks}, leading to the occurrence of a rapidity
gap.

It should be noted, however, that according to this argument the gap
probability is not exactly zero even in CNS exchange
contributions. For example, the squared $t-$channel diagram also
contributes at $\Theta_{CMS} \sim \pi$, which, according to the above
argument, should lead to a rapidity gap. Conversely, the probability
for emission into the gap vanishes only in the limit of vanishing
gluon momentum. This implies that emission into the gap is possible,
although the corresponding probability is not enhanced by large
logarithms. These arguments imply that one will not be able to
distinguish between CS and CNS exchange contributions on an
event--by--event basis. We do nevertheless expect significant
differences in distributions of observables that are sensitive to QCD
radiation.

\subsection{Simple Monte Carlo Simulations}

The discussion of the previous Subsection indicates that the angular
distribution of gluons in $q q \rightarrow \tilde q_L \tilde q_L g$ events
should be very different for CNS and CS exchange contributions. As a first
step, we want to verify this expectation using an explicit parton level
calculation. Here, as for all our numerical studies, we assume the mSUGRA
\cite{msugra} benchmark scenario SPS1a \cite{sps} ($m_0=100$ GeV,
$m_{1/2}=250$ GeV, $A_0=-100$ GeV, $\tan \beta=10 \Rightarrow m_{\tilde
  q}=560$ GeV) with conserved $R-$parity\footnote{The production dynamics,
  including the pattern of QCD radiation, is not sensitive to $R-$parity
  breaking. However, jets produced in LSP decays would complicate the
  analysis.}. Events are generated using the event generator MadGraph
\cite{Maltoni:2002qb}, using the pre--defined SPS1a table of sparticle masses
and branching ratios. We use the CTEQ5L \cite{cteq} parameterization of the
parton distribution functions; correspondingly the one--loop expression for the
strong gauge coupling with five active flavors is taken, where
$\Lambda_{\rm{QCD}}=142$ MeV. For simplicity we only consider the case $q=u$.
Since the QCD effects we are interested in are flavor blind, it is not
necessary to generate the full set of combinations of initial and final
flavors \cite{bddk1}. We regulate infrared singularities by requiring the
gluon to have a transverse momentum in excess of 20 GeV. This is still soft on
the scale of the hard process ($m_{\tilde q} \simeq 560$ GeV), i.e. the
arguments presented in the previous Subsection should still be valid. We also
require the squarks to have pseudorapidity $|\eta| \leq 3.2$.

\medskip

In the following we plot histograms of the quantity $\Delta$ defined
as:  
\beq \label{deltadef}
\Delta = \frac{\eta_g-(\eta_1+\eta_2)/2}{\vert \eta_1-\eta_2 \vert/2} \, ,
\eeq
where $\eta_g$ is the rapidity of the gluon and $\eta_{1(2)}$ is the
rapidity of the first (second) final state squark. Therefore a gluon
radiated into the rapidity region spanned by the two squarks
($\eta_1<\eta_g<\eta_2$ for $\eta_1<\eta_2$ ) gives rise to
$\Delta<1$, whereas a gluon outside of this region
($\eta_g<\eta_1,\eta_2$ or $\eta_g>\eta_1,\eta_2$) leads to
$\Delta>1$. Recall that the squarks are kept stable in this Section.

\medskip

We generated samples of 10,000 events each with pure squared CS and
CNS exchange contributions. This is a further simplification, since in
reality CS (neutralino) and CNS (gluino) exchange contributions can
interfere; we will come back to this point shortly. We do include
interference between $t-$ and $u-$channel contributions within each
class of events. 

Since we wish to look for rapidity gaps, we require the two squarks in
the final state to be well separated in rapidity:
\eqa \label{sq_cut}
\Delta\eta=\vert \eta_1-\eta_2 \vert \,\ge 3.0 \,.
\eqe

\begin{figure}[h!] 
\begin{center}
\rotatebox{0}{\includegraphics[scale=0.7]{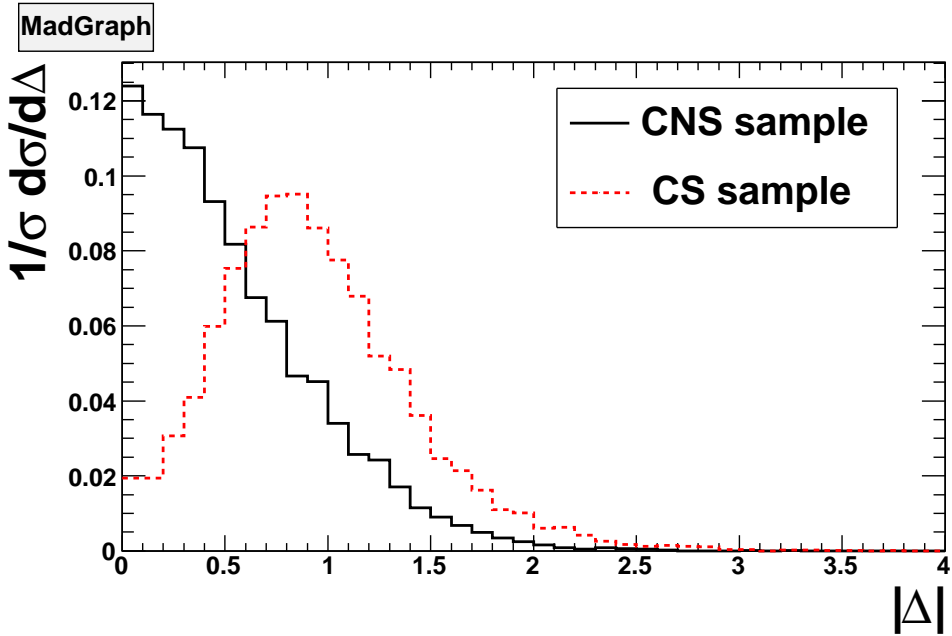}}
\caption{Normalized $\Delta$--distributions of the process
  $uu\rightarrow\tilde u_L\tilde u_L g$ for events generated by MadGraph. Only
  events passing the cut (\ref{sq_cut}) are included in the distribution and
  used for the calculation of the integrated cross section $\sigma$. The black
  (solid) curve denotes the results for the CNS sample (gluino exchange) and
  the red (dashed) curve denotes the results for the CS sample (neutralino)
  exchange. Here, as well as in all subsequent numerical results, we use the
  nominal LHC energy, $\sqrt{s} = 14$ TeV.}
\label{figS1}
\end{center}
\end{figure}

The resulting normalized $\Delta$ distributions, shown in
Fig.~\ref{figS1}, confirm our expectation.  The CNS distribution has
its maximum around $\Delta=0$, i.e. in the middle between the
squarks. It decreases approximately linearly with increasing $\Delta$
values as long as $\Delta \lsim 1.5$. In contrast, the CS distribution
peaks around $\Delta=0.9$, i.e. near the produced squarks, with steep
decreases to both sides. In particular, emission at $\Delta = 0$ is
strongly suppressed by destructive interference between diagrams where
the gluon is emitted off the initial or final state. Note also that,
while suppressed, the CS distribution does not vanish even at $\Delta
= 0$, confirming the caveat we made at the end of the previous
Subsection.

A parton--level simulation is not sufficient to demonstrate the
existence of an experimentally detectable rapidity gap; rather, full
hadron--level simulations are needed. We employ the two generators
PYTHIA 6.4 \cite{Sjostrand:2006za} and Herwig++ \cite{Bahr:2008pv}; in
PYTHIA 6.4, the old shower model, i.e. the virtuality ordered
showering model, is used. These generators can simulate full events,
including parton showers, underlying event and hadronization. However,
we first want to check that they reproduce the exact radiation pattern
shown in Fig.~\ref{figS1}. To this end, we generate $u u \rightarrow
\tilde u_L \tilde u_L$ events with MadGraph. The parton level events
are then interfaced via the SUSY Les Houches Accord
\cite{Skands:2003cj} to the event generators.

This requires that the color flow of each event is specified. We do this
following the procedure of Ref.~\cite{Odagiri:1998ep}, which is also used by
MadGraph. Here, all contributions to a single $2 \rightarrow 2$ scattering
process like $uu\rightarrow\tilde u_L\tilde u_L$ are split up according to
their different color flows, labeled by $i$:
\beq \label{en1}
{\cal M} = \sum_i {\cal M}_i\,.
\eeq
Note that parton shower algorithms generating initial and final state
radiation work with probabilities rather than with quantum field
theoretical amplitudes. However, simply generating a color flow with
probability determined by $\left| {\cal M}_i \right|^2$ could lead to
the wrong total, color--summed probability for generating the given
final state: since different ${\cal M}_i$ can interfere, $\left| {\cal
    M}\right|^2 \neq \sum_i \left| {\cal M}_i \right|^2$. Following
Ref.~\cite{Odagiri:1998ep}, we therefore re--scale the $\left| {\cal
    M}_i \right|^2$:
\beq \label{clf} 
\left|{\cal  M}_{\rm{full},i} \right|^2 = \frac { \left| {\cal M} \right|^2} 
 { \sum_i \left| {\cal M}_i \right|^2} \left| {\cal M}_i \right|^2 \, . 
\eeq
Note that all $\left| {\cal M}_{\rm{full},i} \right|^2$ are positive definite
\cite{Odagiri:1998ep}. Moreover, by construction $\sum_i \left| {\cal
    M}_{\rm{full},i} \right|^2 = \left| {\cal M} \right|^2$. We can
thus generate events with color flow $i$ with probability determined
by $\left| {\cal M}_{\rm{full},i} \right|^2$. 

Nevertheless this method does not properly include the interference
between terms with different color flow, except in the overall
normalization. This is considered acceptable, since these terms are
suppressed by inverse powers of the number of colors $N_c$. We will
follow this practice, and drop all terms that are suppressed by
inverse powers of $N_c$.

To be specific, consider the process $q_i q_j \rightarrow \tilde q_k
\tilde q_l$, where $i,j,k,l \in \{1,2,3\}$ are color indices. Strong
and electroweak $t-$ and $u-$channel diagrams then generate two color
flows. Flow ``1'' is defined through the color tensor $\delta_{il}
\delta_{jk}$, i.e. the color flows from the first quark to the second
squark, and from the second quark to the first squark; flow ``2'' is
defined through the tensor $\delta_{ik} \delta_{jl}$, i.e. the color
flows from the first quark to the first squark, and from the second
quark to the second squark. In the following we drop possible
$s-$channel contributions. These have small matrix elements; at the
LHC they are further suppressed since they require the existence of an
antiquark in the initial state \cite{bddk1}. We then have
\begin{eqnarray} \label{M_i}
{\cal M}_1 &=& {\cal M}_{t,{\rm QCD}} + {\cal M}_{u,{\rm EW}}\,; \nonumber
\\
{\cal M}_2 &=& {\cal M}_{u,{\rm QCD}} + {\cal M}_{t,{\rm EW}}\,.
\end{eqnarray}
Here, ${\cal M}_{t,QCD}$ describes $t-$channel gluino exchange, ${\cal
  M}_{u,{\rm EW}}$ describes the exchange of an electroweak gaugino in
the $u-$channel, and so on.\footnote{All four matrix elements in
  Eq.(\ref{M_i}) are non--zero only if the two quarks in the initial state
  have the same flavor. Otherwise ${\cal M}_{u,{\rm QCD}} = 0$, but ${\cal
  M}_{u,{\rm EW}}$ may still be nonzero if the two quarks belong to an $SU(2)$
  doublet \cite{bddk1}.}

Recall that in Fig.~\ref{figS1} we had considered pure CNS or pure CS
exchange. In order to compare these ``exact'' (leading order) matrix
element results with the results of PYTHIA and HERWIG++, we therefore
``switch off'' either the QCD or the electroweak contributions. To
leading order in $1/N_c$ there is then a one--to--one correspondence
between Feynman diagrams and color flows, i.e. each diagram has a
unique color flow\footnote{This actually holds exactly for the
  electroweak contributions. However, there is a sub--dominant color
  flow in the QCD $t-$ or $u-$channel contributions. For example, the
  $t-$channel QCD amplitude is proportional to $\sum_a \lambda^a_{ki}
  \lambda^a_{lj} = 2 \left( \delta_{kj} \delta_{il} - \delta_{ki}
    \delta_{lj}/3 \right)$. The second term is suppressed by $1/N_c$.}, and
  each color flow only gets contributions from one type of diagram.
Moreover, we require at least one gluon to have $p_T > 20$ GeV. We
generated 30,000 events for each simulation, in order to obtain a
similar number of events after cuts as in the MadGraph simulation.

\begin{figure}[t!] 
\begin{center}
\rotatebox{0}{\includegraphics[scale=0.7]{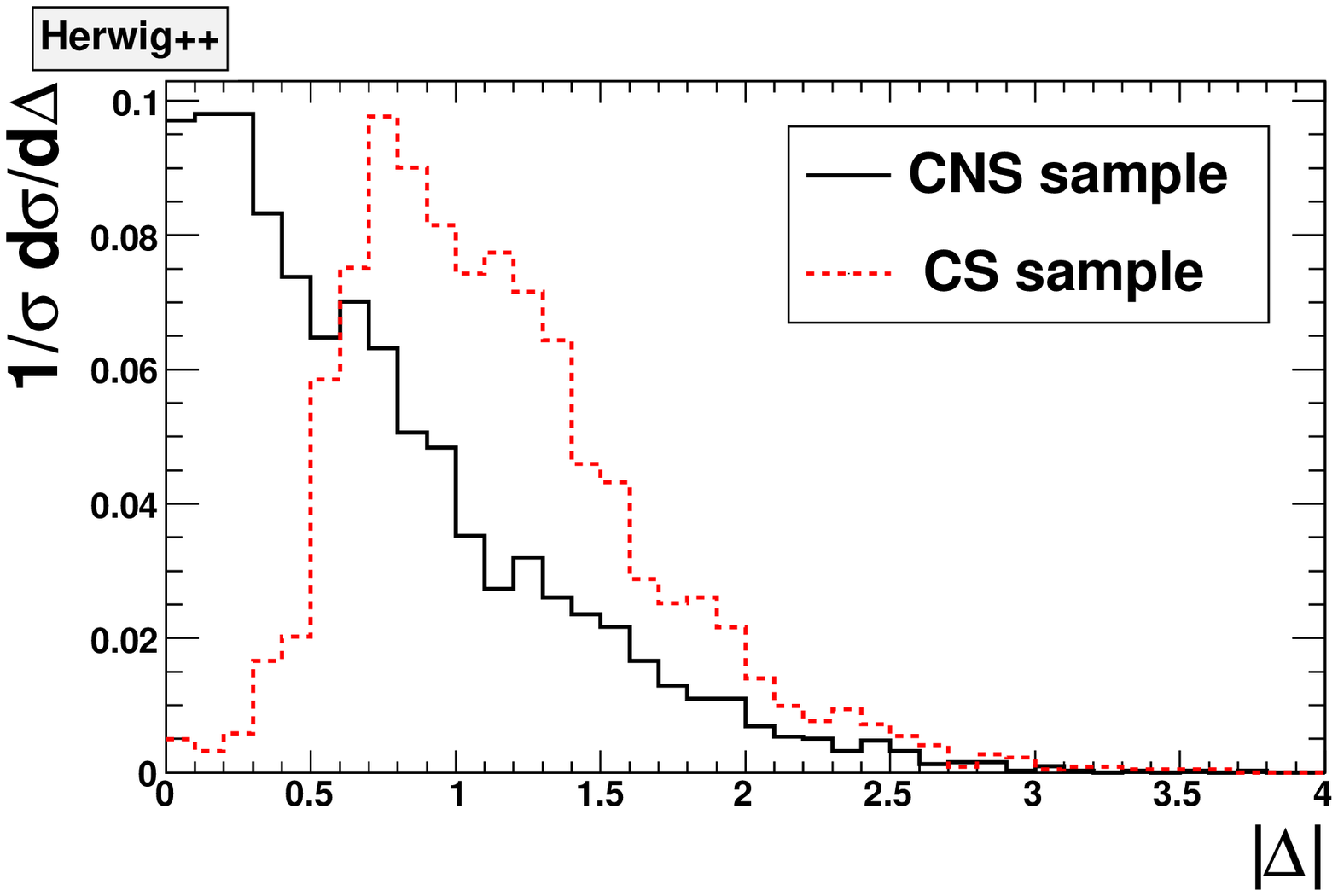}}
\rotatebox{0}{\includegraphics[scale=0.7]{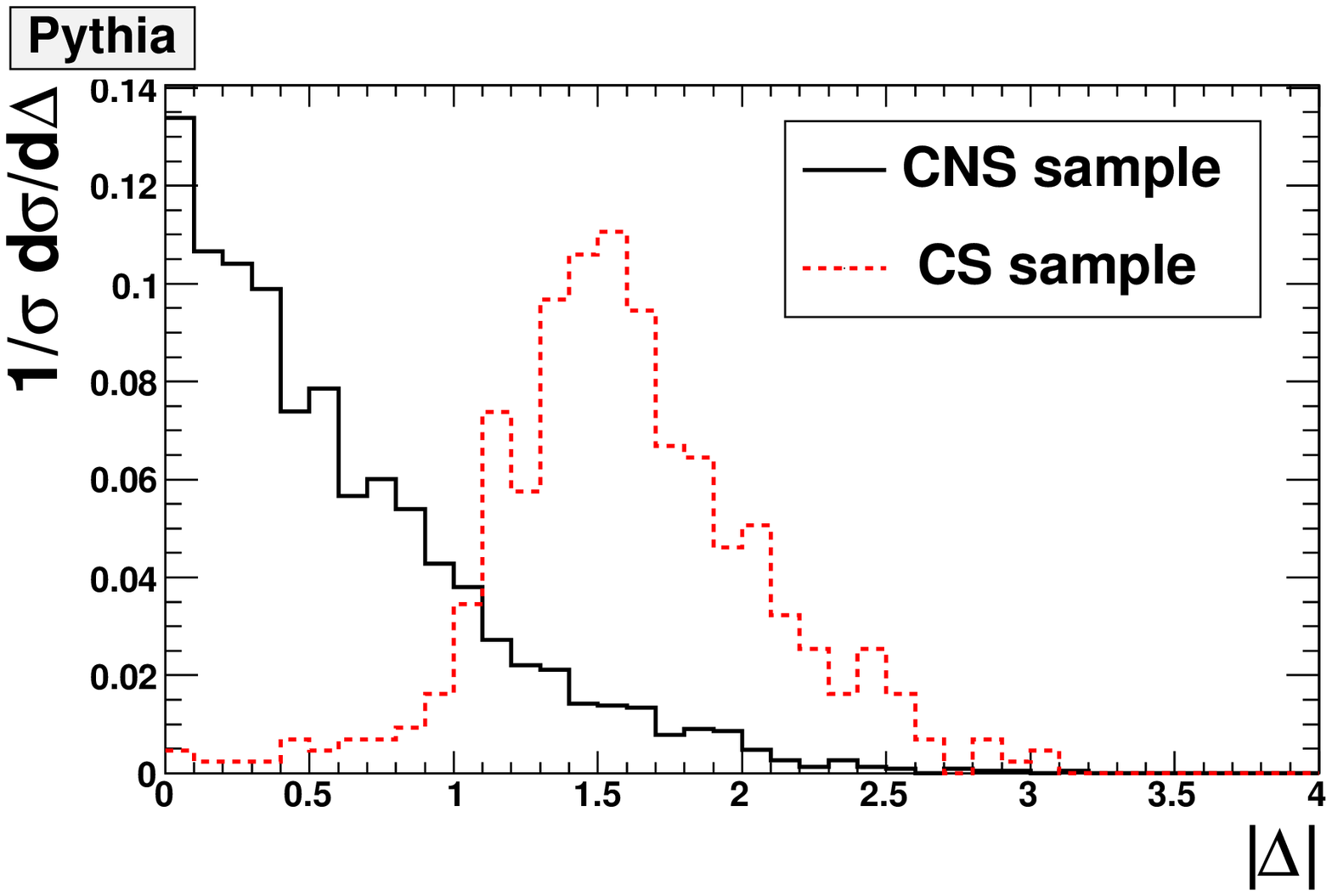}}
\caption{Normalized $\Delta-$distributions for gluon radiation
  simulated with Herwig++ (top) and PYTHIA 6.4 (bottom). Labeling as
  in Fig. \ref{figS1}.}
\label{figS3}
\end{center}
\end{figure}

In Fig.~\ref{figS3} we show the resulting normalized $\Delta$
distributions for the gluon with the largest $p_T$, as predicted by
HERWIG++ (top) and PYTHIA 6.4 (bottom). We see that both event
generators predict large differences between CNS and CS exchange, in
qualitative agreement with the MadGraph prediction of
Fig.~\ref{figS1}.

However, there are also significant discrepancies between the three
predictions. In the CNS case, PYTHIA 6.4 closely reproduces the MadGraph
prediction, whereas HERWIG++ generates a distribution that extends to larger
values of $\Delta$, leading to a less pronounced peak at $\Delta = 0$. In the
CS case, PYTHIA 6.4 generates a distribution that peaks at $\Delta \simeq
1.5$, quite far away from the squarks, with very few gluons populating the
region $\Delta < 0.5$. We should mention here that the ``new shower'' model of
PYTHIA 6.4, which became the default model in the C++ version of PYTHIA
\cite{pythia_c}, failed to predict a gap even in the pure CS exchange case.
This is why we only show results based on the ``old'' showering algorithm; it
is based on virtuality ordering, with angular ordering imposed a posteriori.
In contrast, HERWIG++ predicts the maximum of the distribution to occur at
$\Delta \simeq 0.75$, i.e. between the two squarks, with a very rapid
fall--off towards smaller $\Delta$. As a result, the HERWIG++ prediction for
$\Delta = 0$ also falls below the MadGraph result.

We conclude that both event generators reproduce the gross features of the
normalized MadGraph prediction. Since the former typically generate several
gluons, we do not actually expect exact agreement with the fixed--order
prediction of the latter. The comparison also shows significant differences
between PYTHIA 6.4 and HERWIG++. In fact, the differences between PYTHIA and
Herwig++ become even larger once we consider the unnormalized distributions.
In the pure QCD sample, Herwig++ predicts about 35\% more events with $\delta
\eta \geq 3.0$ to contain at least one gluon with $p_T \geq 20$ GeV.  In the
pure EW sample, this difference between the two generators even amounts to a
factor of seven, with Herwig++ again predicting more gluon radiation. At the
present time these differences should be interpreted as systematic
uncertainties of the predictions. We will therefore continue to show
predictions from both generators. In Sec.~4 we will suggest how this
uncertainty might be reduced using real data.

\subsection{Interference between Color Singlet and Color Non--Singlet
  Exchange}

Before turning to fully realistic simulations, we want to study the
effects of interference of CS (neutralino) and CNS (gluino) exchange
contributions, which we have ignored so far. In fact, at least as far
as the total squark production cross section is concerned, the biggest
effect of electroweak gaugino exchange {\em is} due to interference
with the dominant QCD diagrams \cite{bddk1}. For our benchmark point
SPS1a, this increases the total cross section for the production of
$SU(2)$ doublet, ``$L-$type'' squarks by about 16\%.

To test the effect of the electroweak CS exchange contribution on the
$\Delta$ distribution, we generated 100,000 $u u \rightarrow \tilde
u_L \tilde u_L g$ events with either pure QCD or QCD$+$EW
contributions, again requiring $p_T(g) > 20$ GeV.

\begin{figure}[h!] 
\begin{center}
\rotatebox{0}{\includegraphics[scale=0.7]{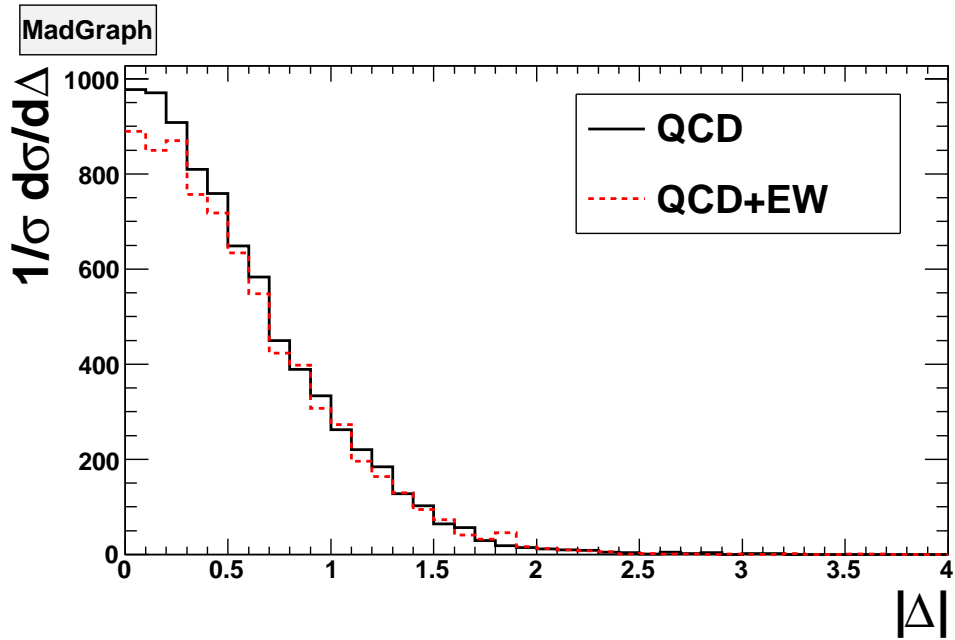}}
\caption{ $\Delta$--distributions for gluon radiation simulated with
  MadGraph. The black (solid) curve denotes the results for the CNS
  sample (pure QCD exchange particles) and the red (dashed) curve
  denotes the results for the CNS+CS sample (QCD and EW exchange
  particles). One has 7962 events after the cut (\ref{sq_cut}) in the
  pure QCD sample and 7506 events in the mixed QCD+EW sample.}
\label{figS5}
\end{center}
\end{figure}

The resulting $\Delta$ distributions are shown in Fig.~\ref{figS5}.
We are showing unnormalized distributions, because it makes the
difference between the two samples more visible and shows the feature
of an efficiency reduction.

The latter is illustrated by the fact that the total number of events with two
squarks having a rapidity--distance larger than three is 7962 for the QCD
sample and 7506 for the QCD+EW sample, so including EW contributions reduces
the efficiency of passing the cut of Eq.(\ref{sq_cut}) by $(6.1 \pm 1.7)\%$
(statistical error only). This can be explained as follows. Requiring a large
$\Delta\eta$ between the two squarks singles out events where the CMS
scattering angle $\theta$ is close to either 0 or $\pi$, and with sizable
squark CMS velocities $\beta$. The latter condition implies that the $t-$ and
$u-$channel propagators depend strongly on $\theta$:
\beq \label{eprop1}
\hat{t} - M^2_{\tilde V} = m^2_{\tilde q} -\frac{\hat{s}} {2} ( 1 - \beta
\cos\theta)  - M^2_{\widetilde V} \, , 
\eeq
where $M_{\tilde V}$ is the mass of the exchanged gaugino and
$\hat{s},\, \hat t$ and $\hat u$ are the partonic Mandelstam
variables; the expression for $u-$channel propagators can be obtained
by the replacement $\cos\theta \rightarrow -\cos\theta$. These
propagators prefer large $\beta |\cos\theta|$; however, $t-$ and
$u-$channel propagators prefer different signs of $\cos\theta$. Thus a
large value of $\Delta\eta$ suppresses the relative importance of the
interference of $t-$ and $u-$channel diagrams. We showed in
Ref.~\cite{bddk1} that the dominant electroweak contributions are
precisely due to such interference terms. Moreover, in pure QCD $t-$
and $u-$channel diagrams interfere destructively, which makes the
peaks at large $|\cos \theta|$ even more pronounced. Therefore the cut of
Eq.(\ref{sq_cut}) reduces the importance of the electroweak
contributions.\footnote{We note that including EW contributions does
  {\em not} reduce the efficiency for $u d \rightarrow \tilde u_L
  \tilde d_L$, since this process does not receive $u-$channel
  contributions in pure QCD.}

For $\Delta \leq 0.2$, the difference between the pure QCD and QCD+EW
predictions amounts to $\sim 10\%$, larger than the integrated
difference of 6\% discussed above. Evidently including EW
contributions slightly decreases the probability to emit a gluon at
small $\Delta$. This can be understood from Eqs.(\ref{M_i}): ${\cal
  M}_1$, where the color flow will lead to a gap if $\theta \simeq
\pi$, receives an electroweak $u-$channel contribution which is in
fact peaked at $\theta = \pi$, where the pure QCD contribution is
minimal; analogous statements hold for ${\cal M}_2$. 

The size of the effect seen in Fig.~\ref{figS5} is smaller than one would
expect from our observation \cite{bddk1} that EW contributions increase the
total cross section for $\tilde u_L \tilde u_L$ production by about 16\%.
There are two reasons for this. First, we argued above that the cut of
Eq.(\ref{sq_cut}) reduces the relative size of these interference terms.
Secondly, in Fig.~\ref{figS5} we are studying the $2 \rightarrow 3$ process $u
u \rightarrow \tilde u_L \tilde u_L g$. Here including the EW contribution
increases the total cross section (with $p_T(g) > 20$ GeV, but without cut on
the squark rapidities) by only 9\%. This last observation indicates that
Fig.~\ref{figS5} understates the importance of EW contributions: in addition
to changing the shape of the $\Delta$ distribution of emitted hard gluons,
they also reduce the probability of emitting a hard gluon {\em anywhere} in
phase space. The importance of this second effect can best be investigated
with the help of event generator programs, which produce $\tilde q \tilde q +
n\,g$ events with (approximately) correct relative weight for any
$n=0,1,2,\dots$. This is the topic of the next Section.

\section{Full Event Simulation}

We now turn to full event simulation, including squark decays, hadronization,
jet reconstruction and the ``underlying event'', however without simulating
the detector. To be specific, we consider rapidity--gap events for squark pair
production at the LHC, where electroweak (EW) contributions at tree level are
included. The production of the first two generations of squarks via $t-$ and
$u-$channel diagrams is taken into account. $s-$channel contributions are
neglected, since they are quite small \cite{bddk1}, and are not expected to
lead to rapidity--gap events. Quark mass effects, the mixing between $SU(2)$
doublet and singlet squarks and flavor mixing effects are neglected. The mass
spectrum and branching ratios of the sparticles are obtained from SPheno
\cite{spheno}.  Analytical expressions for the squared and averaged matrix
elements are given in Ref.~\cite{bddk1}. We implemented the relevant matrix
elements for QCD and as well for EW contributions in a simple parton--level
simulation. Jets are reconstructed via the $k_T$ clustering algorithm of
FastJet \cite{Cacciari:2006sm}. Events are analyzed using the program
package root \cite{Brun:1997pa}.

We saw in Subsec.~2.2 that the angular distributions of the hardest
gluons in PYTHIA 6.4 (with the ``old'' shower algorithm) and Herwig++
roughly agrees with an ``exact'' (leading order) MadGraph calculation.
We therefore continue to use these two generators, hoping that the
difference between their predictions can be used as a measure of the
current theoretical uncertainty. 

Our goal is to find evidence for the different color flows in CS (EW)
and CNS (pure QCD) exchange contributions. In Ref.~\cite{bddk1} we
have shown that production of two $SU(2)$ doublet squarks receives the
largest EW contributions, partly because the $SU(2)$ gauge coupling is
larger than the $U(1)_Y$ gauge coupling. Moreover, most relevant
matrix elements are proportional to the mass of the exchanged gaugino,
and in mSUGRA winos are about two times heavier than the bino. For
example, in SPS1a, the cross section for production of two $SU(2)$
doublet squarks is enhanced by $13\%$ by the EW contributions. If at
least one $SU(2)$ singlet squark is in the final state, EW
contributions are much smaller, leading to an increase of the total
squark pair production cross section of only $4$\%. In order to
enhance the importance of the EW contributions, and hence the rapidity
gap signal, we thus look for cuts that single out the contribution from
the pair production of $SU(2)$ doublet squarks.

This is possible at least for $m_{\tilde g} \gsim m_{\tilde q} >
|M_2|, |M_1|$, which is true for the SPS1a benchmark point we are
considering. In this case $SU(2)$ singlet squarks prefer to decay into
the neutralino with the largest bino component \cite{bbkt}, which is
usually the $\tilde \chi_1^0$ in mSUGRA. In contrast, $SU(2)$ doublet
squarks prefer to decay into charginos and neutralinos dominated by
wino components \cite{bbkt}, which are typically the $\tilde \chi_2^0$
and $\tilde \chi_1^\pm$ in mSUGRA. Since $\tilde \chi_1^0$ is stable
while $\tilde \chi_2^0$ and $\tilde \chi_1^\pm$ can decay
leptonically, the relative contribution of doublet squarks can be
enhanced experimentally by requiring the presence of energetic,
isolated charged leptons, in addition to $\geq 2$ jets and missing
transverse momentum \cite{nojiri}. By requiring the presence of two
leptons with equal charges we single out events where {\em both}
squarks decay semi--leptonically; in contrast, events with two
oppositely charged leptons could come from $\tilde q_R \tilde q_L$
production followed by $\tilde q_L \rightarrow q \tilde \chi_2^0
\rightarrow q \ell^+ \ell^- \tilde \chi_1^0$ decays ($\ell = e,\,
\mu,\,\tau$).\footnote{If $m_{\tilde g} < m_{\tilde q}$ this
  distinction becomes more difficult, because in this case most
  squarks decay into a gluino and a quark; nevertheless, the branching
  ratio for $SU(2)$ doublet squark decays into wino--like states
  remains sizable even in such a case.} Note that we do not veto the
presence of additional charged leptons, i.e. we also accept events
with three or four charged leptons.

In SPS1a, $SU(2)$ doublet squarks $\tilde q_L$ almost always decay
into either $\tilde \chi_1^\pm q'$ or into $\tilde \chi_2^0 q$, with
relative frequency of approximately $2:1$. The electroweak gauginos
$\tilde \chi_2^0$ and $\tilde \chi_1^\pm$ in turn almost always decay
leptonically. Our requirement of two like--sign charged leptons
therefore accepts nearly all $\tilde u_L \tilde u_L$ and $\tilde d_L
\tilde d_L$ pair events, as well as more than half of $\tilde u_L
\tilde d_L$ events (rejecting only events where both $\tilde u_L$ and
$\tilde d_L$ decay into a chargino). Most of the charged leptons
produced in $\tilde \chi_2^0$ and $\tilde \chi_1^\pm$ decays are $\tau$
leptons. For the sake of simplicity, we assume that the $\tau$
detection efficiency is 100$\%$.  The final state from the production
of two $SU(2)$ doublet squarks thus typically contains two or more
$\tau$ leptons, $\ge 2$ jets and missing $E_T$.

Altogether we therefore impose the following cuts. We require that the two 
highest transverse momentum jets satisfy
\begin{equation} \label{etj}
E_T(j_i) \ge 100\,{\rm GeV} \, ; \quad \left|\eta(j_i)\right|
\leq 5.0 \quad (i=1,2).
\end{equation}
We further suppress SM backgrounds by requiring a large amount of
missing transverse energy,
\beq \label{etmiss}
E_T\hspace*{-4.5mm}/ \hspace*{4.5mm} \geq 100 \, {\rm GeV}.
\eeq
Squark pair events containing at least one $SU(2)$ singlet squark are
suppressed by requiring the existence of two like--sign charged
leptons, with
\beq \label{ptl}
p_T(\ell_i) \geq 5 \, {\rm GeV}\,; \quad \left| \eta(\ell_i) \right| \leq
2.4 \quad (i=1,2).
\eeq

In order to be able to define a meaningful rapidity gap, the two
leading jets should be well separated in rapidity:
\begin{equation} \label{etaj}
\Delta\eta \equiv \left| \eta(j_1) - \eta(j_2) \right| \ge3.0 \, .
\end{equation}
We have to take into account that the two jets have finite radii. The
gap region is therefore defined as 
\begin{equation}\label{gap_region}
\rm{min}[\eta(j_1),\eta(j_2)] + 0.7 \le \eta \le
\rm{max}[\eta(j_1),\eta(j_2)] -0.7 \, .
\end{equation}
One can expect that most of the particles produced during
hadronization are within the cone of $0.7$ of the corresponding jets
\cite{Bjorken:1992mj}. 

Since we want to avoid ``event pile--up'', i.e. multiple $pp$ interactions
during the same bunch crossing, we assume an integrated luminosity of 40
fb$^{-1}$ at $\sqrt{s} = 14$ TeV. Using LO cross sections \cite{bddk1}, this
corresponds to about $484,000$ squark pair events for pure QCD, about
$121,000$ of which contain two $SU(2)$ doublet squarks; these numbers increase
to $502,000$ (all squark pairs) and $140,000$ (only $SU(2)$ doublets) events
once electroweak contributions are included.

In our simulation we generated squark pair events, with sample sizes
equal to the expected numbers of events given above. After applying
all cuts, in the Herwig++ simulation, 3574 and 4032 events remain in
the QCD and QCD+EW samples, respectively. In the PYTHIA 6.4
simulation, 3271 QCD and 3723 QCD+EW events are retained. The overall
efficiency of our cuts relative to the entire squark pair sample is
therefore less than 1\%. Of course, this is partly desired. Focusing
on the production of two $SU(2)$ doublets reduces the number of events
by a factor of about four. Moreover, we saw above that requiring two
like--sign leptons removes nearly half of all
 $\tilde d_L \tilde u_L$; this is a pity, since in this channel
the EW contributions are relatively most important \cite{bddk1}. This
reduces the event samples by another factor of about 1.3. 

Since most ``primary'' jets from squark decay are quite central, the
cut of Eq.(\ref{etaj}) reduces the event samples by another factor of about
15. We note that the efficiency of this cut is higher than that of the
analogous cut of Eq.(\ref{sq_cut}) applied to the squarks: since squark
decays release a large amount of energy, the jets resulting from these
decays can be further apart in rapidity than the decaying squarks, if
the quarks are emitted in a direction close to the flight direction of
their squark parents. Since the cut of Eq.(\ref{etaj}) favors such
configurations, in most of the accepted events, squark decay will not
lead to an additional acceleration of the color charge over a large
angle. The discussion of the previous Subsection regarding the angular
region covered by QCD radiation should therefore remain qualitatively
correct. Finally, almost half of the remaining events are removed by
the kinematical cuts of Eqs.(\ref{etj}), (\ref{etmiss}) and (\ref{ptl}). We
note that the total cut efficiency for events containing two $SU(2)$
doublet squarks is nearly the same for the pure QCD and QCD+EW samples
after summing over all channels.

\begin{figure}[h!]
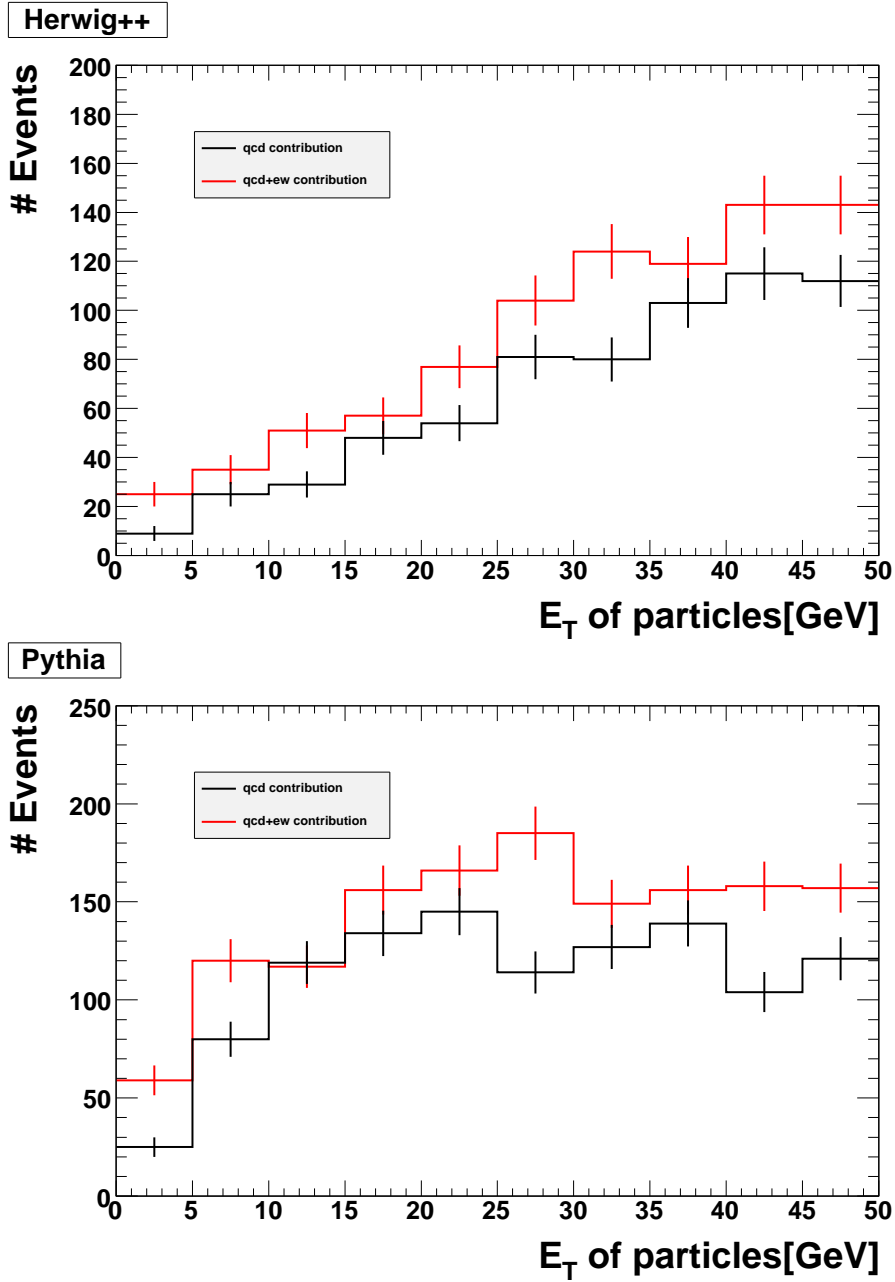
 
\begin{center}
\rotatebox{270}{\includegraphics[scale=0.5]{plot4h.epsi}}
\rotatebox{270}{\includegraphics[scale=0.5]{plot4p.epsi}}
\caption{Transverse energy in the rapidity--gap region
  (\ref{gap_region}) as predicted by full event simulations of squark
  pair production using HERWIG++ (top) and PYTHIA 6.4 (bottom).  Black
  histograms show pure QCD results, while the red (gray) histograms
  include electroweak contributions. The errors are statistical only.}
\label{fig1}
\end{center}
\end{figure}

Our first attempt to isolate ``rapidity gap events'' uses a completely
inclusive quantity. We define $E_{T,{\rm particles}}^{\rm{gap}}$ as the total
transverse energy deposited in the gap region defined in
Eq.(\ref{gap_region}); this is computed from all photons and hadrons in the
event (after hadronization and decay of unstable hadrons), but does {\em not}
include the leptons produced in $\tilde \chi^0$ and $\tilde\chi^\pm$ decays.
The distribution of $E_{T,{\rm particles}}^{\rm{gap}}$ is shown in
Fig.~\ref{fig1} for Herwig++ (top) and PYTHIA~6.4 (bottom). In this and all
following figures, black and red histograms denote pure QCD and QCD+EW
predictions, respectively. We also show the statistical error for each bin.

We note that including EW contributions increases the number of
events, although in most bins this effect is statistically not very
significant. However, in the first bin, where the total $E_T$ is less
than 5 GeV, the inclusion of these CS exchange contributions increases
the number of events by a factor of $2.8 \pm 1.1$ and $2.36 \pm 0.56$
in the Herwig++ and PYTHIA 6.4 simulations, respectively. This
indicates that CS exchange does lead to ``gap'' events where little or
no energy is deposited between the two hard jets.

However, only 0.6\% (1.6\%) of all events passing our cuts in Herwig++
(PYTHIA 6.4) are true gap events in this sense. Evidently PYTHIA
predicts many more such gap events. This may partly be due to the
differences in showering off CS exchange events seen in
Fig.~\ref{figS3}. However, PYTHIA also predicts many more gap events
in a pure QCD (CNS exchange) simulation. We remind the reader that
PYTHIA predicted significantly fewer events to contain a hard gluon
even in a pure SUSY QCD sample. Moreover, the two event generators not
only use different parton shower algorithms; their modeling of
hadronization, and of the underlying event, also differs. Overall, the
difference between the two generators is as large as the effect from
the CS events: PYTHIA 6.4 without CS exchange contributions predicts
almost exactly the same number of events in the first bin as Herwig++
with CS exchange. PYTHIA 6.4 also predicts a $E_{T,{\rm
    particles}}^{\rm{gap}}$ distribution which is quite flat beyond 20
GeV, whereas the distribution predicted by Herwig++ flattens out only
at about 40 GeV. One might thus be able to use the higher bins, where
the effect of the CS exchange contributions is not very sizable, to
decide which generator describes the data better, or to tune the Monte
Carlo generators to the data. This should reduce the difference
between the two predictions.

The results of Fig.~\ref{fig1} only include contributions from squark
pair production. Non--su\-per\-sym\-met\-ric backgrounds are
negligible after our basic cuts of Eqs.(\ref{etj})--(\ref{etaj}). However,
other supersymmetric final states may contribute; in the present
context they have to be considered as background. Owing to the large
cross section, events containing at least one gluino in the final
state are of particular concern.

In order to check this, we generated all hard SUSY $2 \rightarrow 2$
processes with at least one gluino in the final state using
PYTHIA. These processes add up to a cross section of about 24 pb for
SPS1a, which exceeds the squark pair production cross section by a
factor of two. With only the basic cuts of Eqs.(\ref{etj})--(\ref{etaj}),
these additional contributions more than double the entries of the
first $E_T$ bin in Fig.~\ref{fig1}, compared to the result for CNS
squark pair production. This would reduce the significance
noticeably. However, gluino events practically always contain at least
one additional hard jet from $\tilde g \rightarrow \bar q + \tilde q,\
q + \tilde q^*$ decays. This SUSY background can thus be suppressed
efficiently, with negligible loss of signal, by vetoing the presence
of a third energetic jet. In the following we therefore continue to
focus on squark pair events.

\begin{figure}[h!] 
\begin{center}
\rotatebox{270}{\includegraphics[scale=0.5]{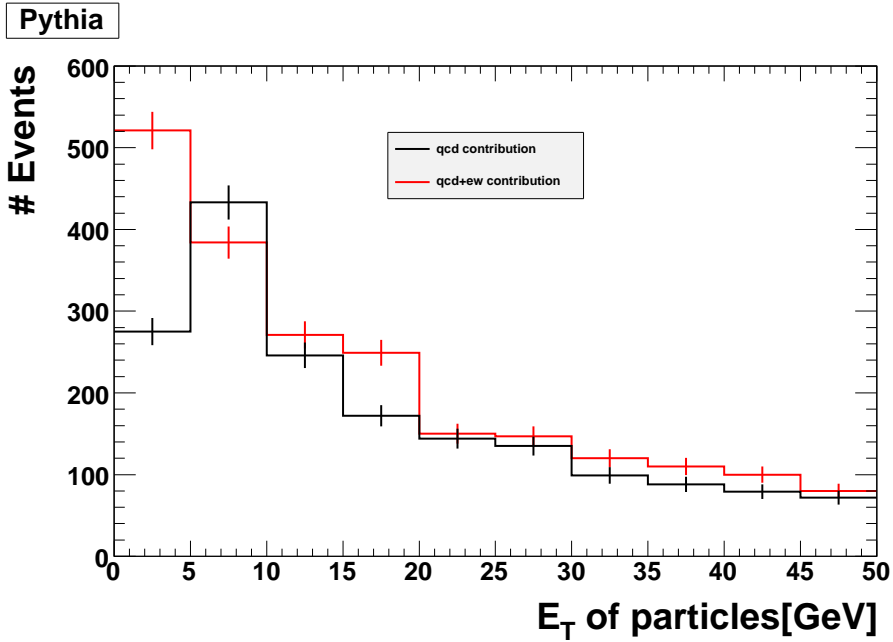}}
\caption{Same as Fig.~\ref{fig1}, bottom frame (PYTHIA 6.4
  prediction), except that the underlying event has been switched off.}
\label{nompi}
\end{center}
\end{figure}

One reason for the small number of true gap events, and the
correspondingly large statistical error, is the energy deposition of
the ``underlying event'', which describes the beam remnants. These
``spectator partons'' do not participate in the primary partonic
squark pair production reaction, but may interact via semi--hard QCD
reactions \cite{zijl}. The underlying event can thus deposit a
significant amount of transverse momentum, with little or no phase
space correlation with the primary jets. The importance of this
component of the total $E_T$ flow is illustrated in Fig.~\ref{nompi},
which shows the PYTHIA~6.4 prediction for the total $E_{T,{\rm
    particles}}^{\rm{gap}}$ distribution with the underlying event
switched off. Clearly the first few bins now contain many more events
than in Fig.~\ref{fig1}.  For example, we now have 521 (278) entries
in the first $5$ GeV bin for QCD+EW (QCD) simulation, as compared to
59 (25) in Fig.~\ref{fig1}. Recall that including EW contributions
increased the PYTHIA~6.4 event sample after cuts by only about 450
events; evidently about 50\% of these additional events would be true
gap events, if the underlying event were absent. The underlying event
in PYTHIA 6.4 thus leads to a gap ``survival probability''
\cite{ppjets} of $\sim 10\%$ at the LHC.

\begin{figure}[h!]
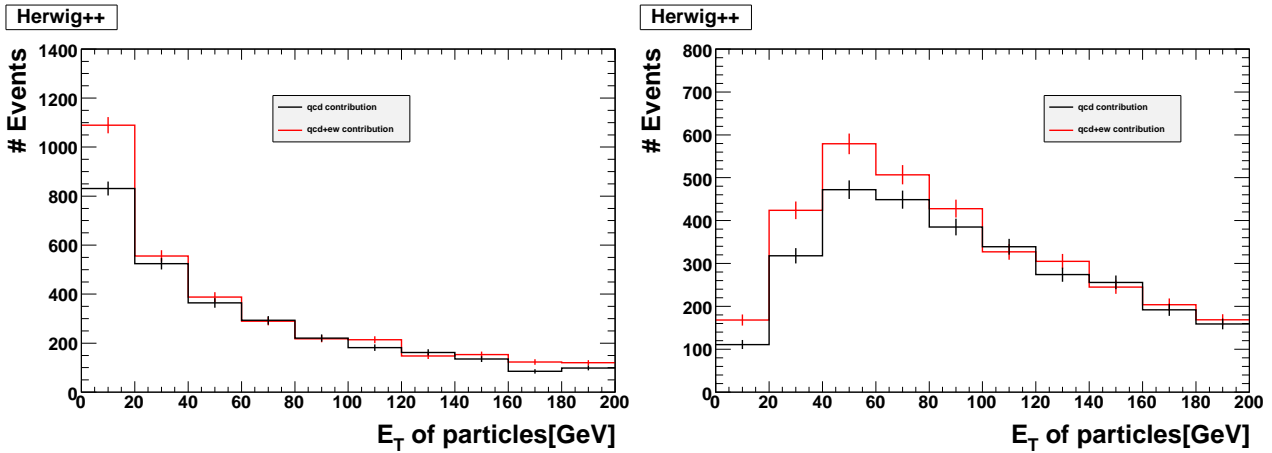

\begin{center}
\rotatebox{270}{\includegraphics[scale=0.35]{plot4hlarge_nompi.epsi} }
\rotatebox{270}{\includegraphics[scale=0.35]{plot4hlarge.epsi}} 
\caption{Same as Fig.~\ref{fig1}, top frame (Herwig++ prediction),
    except that the $x-$axis has been extended to 200 GeV; moreover,
    in the left frame the underlying event has been switched off.}
  \label{twoplots}
  \end{center}
\end{figure} 

The left frame of Fig.~\ref{twoplots} shows analogous results for the
Herwig++ simulations, where we have extended the $x-$axis to 200 GeV.
In this case, only about 50\% of the additional events due to EW
interactions appear in the first bin, even though the bin width has
been increased to 20 GeV; we again observe a smaller effect due to CS
exchange than in the PYTHIA simulation. Note that also the pure QCD
prediction now peaks in the first (wide) bin. In Herwig++ the
underlying event typically deposits about 45 GeV of transverse energy
in the gap. Switching this effect on thus leads to distributions with
broad peaks around 50 GeV, as shown in the right frame of
Fig.~\ref{twoplots}.

Of course, the $E_T$ deposited by the underlying event differs from
event to event. Therefore some true gap events do remain even if the
underlying event has been switched on, as we saw above. In fact, the
effect of the underlying event is somewhat larger in the pure QCD
simulations than in the QCD+EW case. In the Herwig++ simulation of
Fig.~\ref{twoplots}, the underlying event reduces the number of events
in the first bin for the pure QCD (QCD+EW) simulation by about a
factor of eight (six). Figs.~\ref{fig1} and \ref{nompi} show a similar
effect for PYTHIA 6.4: the underlying event reduces the number of
events in the first (5 GeV wide) bin by a factor of about ten (eight)
for the pure QCD (QCD+EW) case. This is not implausible, since the
beam remnants are color connected to the partons participating in
squark pair production, leading to some correlation between the
``hard'' and ``underlying'' parts of the total event.

The significance of the enhanced rapidity gap signal due to EW effects depends
on how exactly the comparison is done. Let us focus on Herwig++ predictions
for definiteness; PYTHIA predicts somewhat larger significances. Simply
looking at the first bin of Fig.~5 gives a statistical significance of about
2.5$\sigma$ given the current MC statistics. However, in an experiment one
would compare the actual event number with a prediction of pure QCD. Assuming
this prediction indeed remained at 9 events with negligible MC error, and the
experiment indeed counted 25 events, the statistical significance would be
$16/\sqrt{9}=5.3\sigma$!\footnote{The Poisson probability of finding 25 or
  more events if 9 are expected is only $8.7 \times 10^{-6}$.}  On the other
hand, the absolute normalization of the curves in Figs.~5 and 7 cannot really
be trusted, since they rely on LO cross sections. If we want to analyze the
change of the shape of the distribution, the larger $E_T$ range depicted in
Fig.~7 is more appropriate. According to the right frame of that figure, pure
QCD and QCD+EW predictions agree to 2\% when all bins with $E_T \geq 100$ GeV
are summed. The sum up to 100 GeV then differs by 5.3$\sigma$ according to
current MC statistics; this could become 8 standard deviations with infinite
MC statistics, assuming the central values remain the same.

\begin{figure}[h!]
\begin{center}
\rotatebox{270}{\includegraphics[scale=0.5]{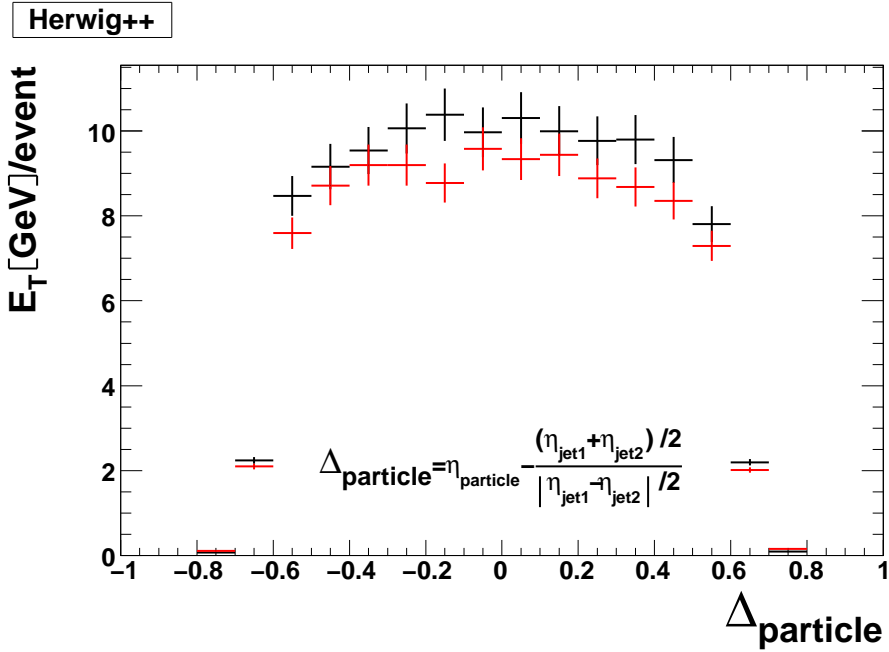}}
\caption{The average amount of deposited transverse energy per bin as
  a function of $\Delta$ as predicted by Herwig++. The black (red or
  grey) points denote the results for the pure QCD (QCD+EW)
  sample. The vertical error bars indicate the statistical
  uncertainty, while the horizontal bars show the bins.}
  \label{etflow}
  \end{center}
\end{figure} 

We also studied the average $E_T$ flow as a function of the scaled
rapidity variable $\Delta$, defined analogously to
Eq.(\ref{deltadef}):
\beq \label{delta_n}
\Delta = \frac{\eta - \left[\eta(j_1) + \eta(j_2) \right]/2} {\left|
    \eta(j_1) - \eta(j_2) \right| /2} \, .
\eeq
The $E_T$ flow in the gap region as predicted by Herwig++ is shown
in Fig.~\ref{etflow}. Recall that the rapidity gap region as defined in
Eq.(\ref{gap_region}) is somewhat smaller than the region $|\Delta| <
1$; this explains the sharp drop of the average deposited $E_T$ for
$|\Delta| > 0.6$. We see that including EW, CS exchange contributions
reduces the average deposited $E_T$ by about 8\%, or 8.5 GeV per unit of
$\Delta$, around $\Delta = 0$. The error bars in Fig.~\ref{etflow} show
that this effect is statistically quite significant. However, given
the large modeling uncertainty, and the sizable energy measurement
errors, the size of the effect is presumably too low to be physically
significant. 

Predicting the total transverse energy flow is difficult, since this
observable is strongly affected by semi-- and non--perturbative
effects. We could try to reduce the importance of these effects by
focusing on charged particles whose $p_T$ exceeds a few GeV
\cite{Dokshitzer:1991he}. Instead we go one step further and discuss
the occurrence of relatively soft ``minijets'' in the ``gap region''
defined in Eq.(\ref{gap_region}).

\begin{figure}[h!]
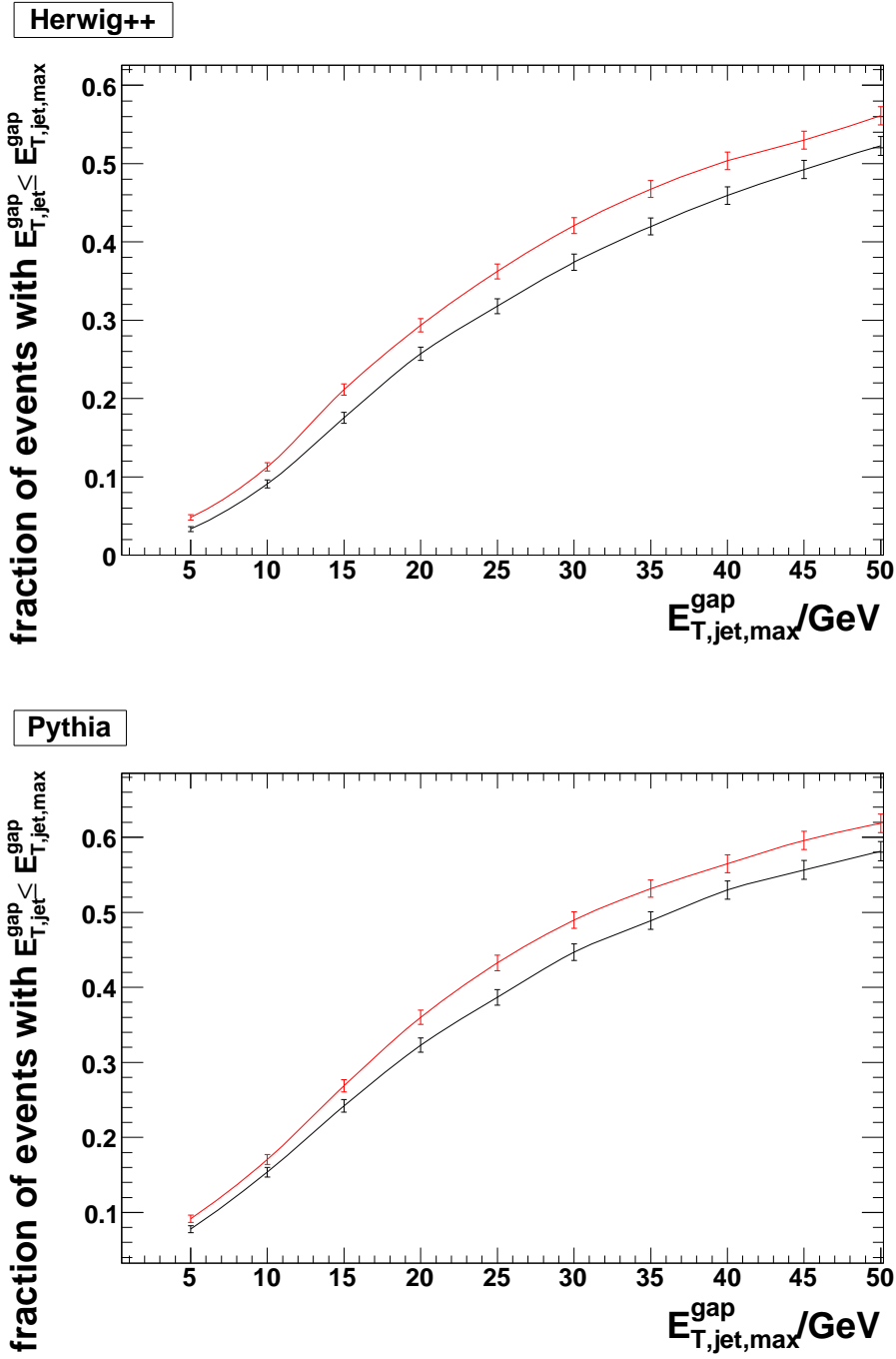
 
\begin{center}
\rotatebox{270}{\includegraphics[scale=0.5]{plot3h.epsi}}\\
\vspace*{8mm}
\rotatebox{270}{\includegraphics[scale=0.5]{plot3p.epsi}}
\caption{Fraction of squark pair events passing a minijet veto in the
  rapidity--gap region (\ref{gap_region}), as predicted using full
  event simulations using Herwig++ (top) and PYTHIA~6.4 (bottom).  The
  black curve is for the pure QCD sample, and the red (gray) curve for
  the QCD+EW sample.}
\label{fig4}
\end{center}
\end{figure}

Fig.~\ref{fig4} shows the fraction of events where the energy
$E_{T,\rm{jet}}^{\rm{gap}}$ of the most energetic jet in the gap region
(\ref{gap_region}) is less than the value $E_{T,\rm{jet,max}}^{\rm{gap}}$
displayed on the $x-$axis. Since Fig.~\ref{fig4} shows event {\em fractions},
all curves asymptotically approach 1 at large $E_{T,\rm{jet,max}}^{\rm{gap}}$.
We assume that jets with transverse energy above $E_{T,{\rm thresh}} = 5$ GeV
can be reconstructed. If the true threshold is higher, the curves should
simply be replaced by constants for $E_{T,\rm{jet,max}}^{\rm{gap}} \leq
E_{T,{\rm thresh}}$. We expect that the underlying event by itself generates
few, if any, reconstructable jets. The results here are nevertheless still not
quite immune to non--perturbative effects, since reconstructed jets may also
contain a few particles stemming from the underlying event. Note also that a
jet whose axis lies in the gap region might contain (mostly quite soft)
particles that lie outside of this region. Conversely, even though we use a
cluster algorithm, where by definition each particle belongs to some jet, some
particles in the gap region might be assigned to a jet whose axis lies outside
this region.

We see that PYTHIA~6.4 (lower frame) predicts more events without jet
in the gap region than Herwig++ (upper frame). This is consistent with
Fig.~\ref{fig1}, where PYTHIA also predicted more events with little
or no energy deposited in the gap region. It also conforms with our
observation at the end of Subsec.~2.2 that PYTHIA tends to generate
fewer hard gluons than Herwig++ does. In fact, for $E_{T,{\rm
    jet,max}}^{\rm gap} = 20$ GeV, which corresponds to the cut
$p_T(g) > 20$ GeV employed in Subsec.~2.2, we again find a ratio of
about 1.3 between the PYTHIA and Herwig++ predictions for the pure
SUSY QCD case. 

Moreover, both PYTHIA~6.4 and Herwig++ predict a significant increase
of the fraction of events without jet in the gap region once EW, CS
exchange contributions are included; the effect is statistically most
significant for $E_{T,\rm{jet,max}}^{\rm{gap}} \sim 20$ to 40
GeV. Here both generators predict an increase of the fraction of
events without (sufficiently hard) jet in the gap by about
0.05. Taking a threshold energy of 30 GeV as an example, Herwig++
predicts about 1,570 out of the total of 4,032 events without jet in
the gap once EW effects are included. This should allow to measure the
fraction of events without jet in the gap to an accuracy of about
0.01; a shift of the jet--in--the gap fraction by 0.05 thus
corresponds to a change by about five standard deviations (statistical
error only). 

Unfortunately a pure SUSY QCD PYTHIA simulation leads to a fraction of
events without jet of 0.45 for the same threshold energy, which is
{\em higher} than the Herwig++ prediction {\em including} EW
contributions. Clearly these large discrepancies between the two MC
generators have to be resolved before reliable conclusions about the
color flow in squark pair events can be drawn.

\section{Tuning MC Generators with SM QCD}

We saw in the last Section that the systematic differences between PYTHIA and
Herwig++ are larger than the physical differences between the QCD and QCD+EW
data samples. In this Section, we demonstrate that PYTHIA and Herwig++ make
similarly different predictions for standard QCD di--jet events. These pure
QCD events can thus be used to tune the Monte Carlo generators. Here
``tuning'' refers to both the setting of parameters (shower scales etc.), and
to details of the algorithms used to describe parton showers and the
underlying event.

\vspace*{7mm}
\begin{figure}[h!] 
\begin{center}
\rotatebox{270}{\includegraphics[scale=0.5]{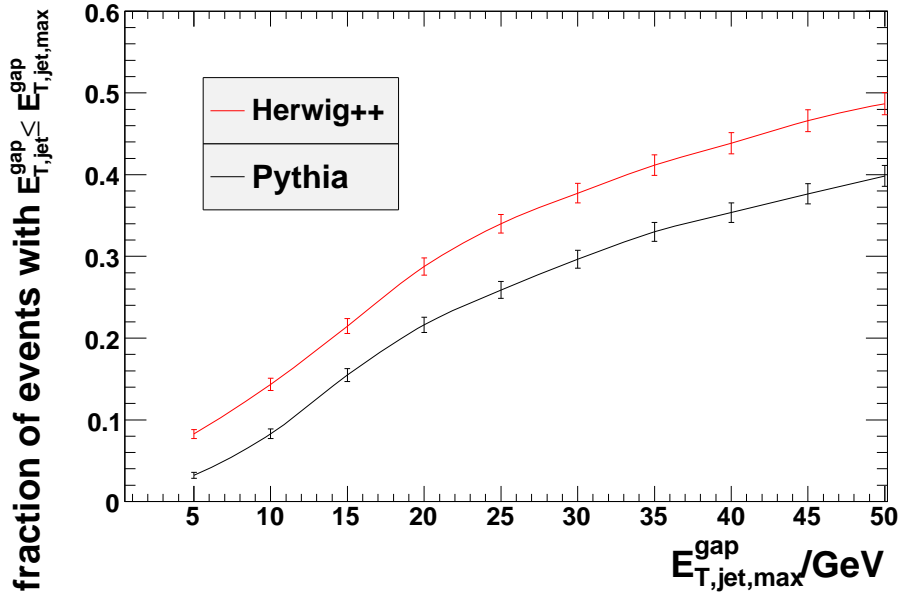}}
\caption{Fraction of events without hard jet in the rapidity--gap
  region for the pure QCD $2 \rightarrow 2$ processes as function of
  the jet energy threshold. The effects of the underlying event,
  parton shower and hadronization are included. The black (red) curve
  denotes the results for PYTHIA (Herwig++).}
\label{fig9}
\end{center}
\end{figure}

To illustrate this, we generated standard QCD di--jet events, where we
require the transverse momentum of the jets to exceed 500 GeV, so that
the kinematics and the relevant Bjorken$-x$ values are comparable to
the squark pair events discussed in the two previous Sections. We
include all standard QCD $2 \rightarrow 2$ processes, i.e. we do not
demand that the final state only consists of quarks. The color
structure may therefore be different than in SUSY QCD squark pair
events. However, the large transverse momentum, as well as the
required large rapidity gap, require quite large Bjorken$-x$ values.
This enhances the contribution from $qq \rightarrow qq$ scattering,
which has exactly the same color structure as the $qq \rightarrow
\tilde q \tilde q$ events in SUSY QCD.

The results are shown in Fig.~\ref{fig9}, where we again plot the fraction of
events without hard jet in the gap region, as function of the threshold
energy. We see that PYTHIA again predicts less radiation, i.e. more events
without hard jet in the gap. For example, for a threshold energy of 20 GeV, we
find the by now familiar ratio of about 1.3 between the predictions of
Herwig++ and PYTHIA. This difference is statistically highly significant.
Quite likely neither of the two predictions presented here will turn out to be
exactly correct. However, given their many successes in describing quite
intricate features of hadronic events, we are confident that soon after real
data for this jet--in--the--gap fraction become available, updated versions of
both PYTHIA and Herwig++ will be produced that are able to reproduce these
data. It seems likely that these advanced versions will then also make very
similar predictions for observables that are sensitive to the color flow in
squark pair events.

We conclude this Section with two remarks. First, the fraction of events
without a hard jet in the gap is significantly smaller in Fig.~\ref{fig9} than
for the case of squark pair production shown in Fig.~\ref{fig4}. This is
mostly due to the remaining contribution containing at least one gluon in the
hard $2 \rightarrow 2$ process, in either the initial or final state. Note
that gluon emission off gluons is enhanced by a color factor of 9/4 relative
to gluon emission off quarks. The presence of a gluon in the hard processes
thus increases the probability of emitting a rather energetic gluon in the
parton shower.

Secondly, the pure QCD prediction for the jet--in--the--gap cross
section at the LHC has recently been calculated \cite{forshaw}. Their
final result agrees quite well with predictions of Herwig++; no
comparison with PYTHIA is provided. However, this agreement is partly
accidental. On the one hand, the theoretical calculation is asymptotic
in the sense that energy conservation is not enforced. On the other
hand, it includes true quantum effects which cannot be fully included
in a conventional MC generator that works with probabilities rather
than amplitudes. We conclude that, while such comparisons between MC
generators and advanced parton level calculations are certainly
useful and important, they cannot replace a comparison with real data.


\section{Summary and Conclusions}

In this paper we analyzed rapidity gap events in squark pair
production, where QCD as well as electroweak (EW) contributions were
taken into account. The different flow of color charges in the two
cases led us to expect that events with EW gaugino exchange should
have a higher probability to show a rapidity gap between the produced
squarks, where little or no energy is deposited, since radiation into
the gap region should be suppressed. An explicit fixed--order
calculation of the $2\rightarrow3$ process $uu\rightarrow \tilde u_L
\tilde u_L g$ using MadGraph confirmed this expectation:
color--singlet exchange leads to a very different rapidity
distribution of the emitted gluon than color--octet exchange.

We also compared this fixed--order calculation to the distribution of the
hardest gluon produced by the event generators PYTHIA~6.4 and Herwig++.  We
confirmed that both generators predict rapidity distributions of hard QCD
radiation which are sensitive to the color structure of the event. In the case
of color non--singlet exchange we found satisfactory agreement of the
normalized distributions between both MC generators and MadGraph. However, the
normalized distributions in the case of pure color--singlet exchange are
noticeably different between PYTHIA, Herwig++ and MadGraph. Moreover, the MC
generators predict quite different probabilities for emitting a gluon with
transverse momentum above our cut--off of 20 GeV.

We next included interference between QCD and EW contributions.  We showed
that the EW $t-$channel diagrams have the same color flow (to leading order in
the number of colors) as QCD $u-$channel diagrams, and vice versa. This allows
to properly include interference between QCD and EW contributions, which
greatly increases the importance of the latter; note that MC generators
require a unique color flow be assigned to each event.

In Sec.~3 we performed a full simulation, including the underlying
event, the full shower algorithm, squark decay, and the hadronization
of all partons. The mSUGRA point SPS1a was employed as a benchmark
scenario. We focused on events containing two charged leptons ($e, \,
\mu$ or $\tau$) with equal charge in addition to at least two jets;
this enhances the contribution from $SU(2)$ doublet squarks, where EW
contributions are much more important than for the production of
$SU(2)$ singlets.

We found two observables yielding distinct results between pure SUSY QCD
simulations and those including EW contributions: the transverse energy of all
particles in the ``rapidity gap'' region between the jets resulting from
squark decay, and the fraction of events not containing an additional jet in
the gap region with $E_T$ above a certain threshold. In the case of the total
transverse energy in the gap, we found the most distinct difference between
the QCD and QCD+EW predictions in the very first bin, i.e. for the lowest
$E_T$ values. Not surprisingly, this difference is smeared out considerably
when the underlying event is included.  The second observable, the
jet--in--the--gap cross section, is less sensitive to non-- or
semi--perturbative effects. Here we found statistically significant
differences between the QCD+EW and QCD predictions for a range jet thresholds
up to 50 GeV.

Unfortunately, for both observables the systematic difference between
PYTHIA and Herwig++ currently make it impossible to make clear--cut
predictions. In particular, PYTHIA's pure QCD prediction is usually
comparable to, or even larger than, the QCD+EW prediction of
Herwig++. In Sec.~4 we showed that a similar difference between the
predictions of the two generators also exists for the
jet--in--the--gap cross section in standard QCD. This indicates that
such standard QCD events can be used during the earliest phases of LHC
running to improve the event generators, hopefully to the extent that
the difference between their predictions becomes significantly smaller
than the effect from the EW contributions. 

We also noted that Herwig++ seems to reproduce sophisticated perturbative QCD
calculations of the SM jet--in--the--gap cross section fairly well. If we
conservatively assume that the Herwig++ predictions, which lie below those of
PYTHIA, are correct also for the SUSY case, the assumed integrated luminosity
of 40 fb$^{-1}$ will suffice to show evidence for color singlet exchange
contributions from the total $E_T$ distribution in the gap at the level of
about 5 to 8 standard deviations; see the discussion of Figs.~5 and 7.
Similarly, for a threshold energy of 30 GeV for the jet in the gap, the
fraction of events without a jet in the gap changes by about five standard
deviations when EW effects are included, see Fig.~9.

Our analyses were based on standard QCD and EW perturbation theory,
augmented by the bells and whistles of the MC generators; it resembles
the work of ref.\cite{chehime} for standard jet production. In
particular, we did not include hard QCD color singlet exchange
contributions. In standard QCD this becomes possible once several
gluons are exchanged. In ref.\cite{enberg} a (non--asymptotic)
version of such ``BFKL dynamics'' \cite{bfkl} (aka perturbative
Pomeron exchange) has been shown to be able to reproduce early
Tevatron data \cite{tev_gap} on rapidity gaps between jets. In the
case at hand one would have to consider exchange of a ``perturbative
Pomeronino'', which includes simultaneous gluon and gluino exchange in
the $t-$ or $u-$channel. We are not aware of any study of this kind of
contribution.

Finally, we remind the reader that much larger EW effects are possible
if the ratio of electroweak to strong gaugino masses is increased
beyond the ratio of approximately 1:3 assumed in mSUGRA. We thus
conclude that experimental studies of the color flow in squark pair
events, of the kind sketched in this paper, are well worth the effort
once the existence of squarks has been firmly established.


\subsection*{Acknowledgments}
This work was partially supported by the Helmholtz-Alliance ``Physics at the
Terascale''. SB wants to thank the ``Universit\"atsgesellschaft Bonn --
Freunde, F\"orderer, Alumni e.V.'' and the ``Bonn-Cologne Graduate School of
Physics and Astronomy'' for financial support. JSK wants to thank the
Helmholtz-Alliance for financial support and the University of Bonn and the Bethe Center for hospitality during numerous vists.
HD would like to thank the Aspen
Center for Physics for hospitality, where part of this work was completed.

\begin{appendix}

\end{appendix}

\end{document}